\documentclass[12pt]{iopart}
\usepackage{iopams}
\usepackage{graphicx}
\begin{document}

\title[Two-particle propagator with the help of Hubbard-I
approximation]{Evaluation of the two-particle propagator for Hubbard model
with the help of Hubbard-I approximation}

\author{A.V. Rozhkov and A.L. Rakhmanov}
\address{Institute for Theoretical and Applied Electrodynamics,
Russian Academy of Sciences, ul. Izhorskaya, 13, 125412, Moscow, Russia}

\begin{abstract}
The Hubbard-I approximation is generalized to allow for direct evaluation
of the equal-time anomalous two-electron propagator for Hubbard model on
two-dimensional square lattice. This propagator is compared against the
quantum Monte Carlo data obtained by Aimi and Imada [J. Phys. Soc. Jpn.
{\bf 76}, 113708 (2007)] in the limit of strong electron-electron
interaction. The Hubbard-I predictions are in a good qualitative agreement
with the Monte Carlo results. In particular, $d$-wave correlations decay
as $c r^{-3}$ (``free electron'' behaviour), if separation $r$ exceeds 2-3
lattice constants. However, the Hubbard-I approximation underestimates
coefficient $c$ by a factor of about three. We conclude that the Hubbard-I
approximation, despite its simplicity and artefacts, captures the
qualitative behaviour of the two-particle propagator for the Hubbard model,
at least for moderate values of $r$.
\end{abstract}

\maketitle

\section{Introduction}

\subsection{General outlook}

Consider the Hubbard model's Hamiltonian~\cite{HubI}:
\begin{equation}\label{Ham}
H=H_{\rm kin}+H_{\rm
int}-\mu\sum_{\mathbf{i},\sigma}n_{\mathbf{i}\sigma},
\end{equation}
where
\begin{equation}
\label{Ham1}
H_{\rm kin}
=
-t\sum_{\mathbf{i,a},\sigma}
	c^\dag_{\mathbf{i} \sigma}
	c^{\vphantom{\dagger}}_{\mathbf{i+a}\sigma},
	\quad
H_{\rm int}
=
U\sum_{\mathbf{i}}n_{\mathbf{i}\sigma}n_{\mathbf{i}\bar{\sigma}}.
\end{equation}
Here $c^\dag_{\mathbf{i} \sigma}$ and $c_{\mathbf{i}
\sigma}^{\vphantom{\dagger}}$ are the creation and annihilation operators
for an electron with spin projection $\sigma$ on site $\mathbf{i}$ of the
square lattice, $t$ is the hoping integral, $U$ is the on-site Coulomb
repulsion, $\mathbf{a}$ are vectors connecting the site $\mathbf{i}$ with
its nearest neighbors, $\mu$ is chemical potential,
$n_{\mathbf{i}\sigma}
=
c^\dag_{\mathbf{i} \sigma}
c_{\mathbf{i} \sigma}^{\vphantom{\dagger}}$
is the operator of electron number, and
$\bar{\sigma}$ means \emph{not}-sigma.

This Hamiltonian is commonly used to describe systems of strongly
correlated electrons. Among these, we mention such important examples as
high-Tc superconducting cuprates, manganites, cobaltites, etc. However,
there is no commonly accepted method that allows to solve Hubbard model for
a generic value of doping.

One of the most intriguing problems of the Hubbard model is the question if
the model captures the essential physics of the high-temperature
copper-oxide superconductors. At low interaction one expects that the model
exhibits Kohn-Luttinger superconductivity 
\cite{KL,loh,yanagisawa,yanagisawaII}.
Unfortunately, the critical temperature in such a regime is very small.
Thus, it is necessary to study the Hubbard model at moderate or large
interaction strengths. Such a regime is tractable only when the
concentration of electrons is low
\cite{kagan,kaganII,kaganIII}.
However, the electron concentration for all known copper-oxide
superconductors is close to one electron per unit cell. Little rigorous
knowledge is available for this limit. Different approaches return quite
controversial answers: some Monte Carlo (MC) studies argue against
superconductivity
\cite{zhang-carlson-no-supercond},
or against high-$T_c$ superconductivity \cite{Ai} in Hubbard model, other
numerical works favor existence of the superconducting ground state with
high condensation energy \cite{yes-supercond}.

In the moderate and strong coupling limits there is no guidance from 
analytical studies as well: in this parameter region the majority of
theoretical methods become unreliable. Thus, one is forced to utilize
uncontrollable devices with unknown accuracy.

In such a situation it appears useful to investigate reliability of
uncontrollable schemes. A particularly convenient uncontrollable
approach for the analysis of the Hubbard model is Hubbard-I
approximation~\cite{HubI}. Hubbard-I, being invalid for
$U\lesssim zt$
(where $z$ is the number of the nearest neighbor sites on the lattice),
could be considered as a good first approximation in the case of the strong
correlations $U\gg zt$. Its evident advantage is its simplicity. The method
allows deriving analytical results in many cases or performing numerical
calculations at low computational cost.

The clarity and transparency of the Hubbard-I approximation explains its
popularity among the researchers working in the field of strongly
correlated electronic systems. For us, it is especially important that,
after the discovery of high-Tc superconductors, the method has been applied
to the superconductivity as well. Namely, Hubbard-I approximation is used
for study of superconductivity in the Hubbard or extended Hubbard model
\cite{domanski,mierzII},
in the Hubbard model with phonons
\cite{zielinski,mierz,cebula},
in the Hubbard model with attractive interaction
\cite{caix,caixII}.
A related approach, the two-pole approximation (which, in some respects, is
superior to Hubbard-I approximation), and its generalization are applied to
study possible superconductivity of the usual Hubbard model
\cite{beenen}
and of the multi-band Hubbard model \cite{gomes}. 
Similar techniques are used in \cite{plakida,ovchinnikov_valkov}.
Systematic expansion in orders of $t/U$ is developed in \cite{zaitsev}.
It contains the Hubbard-I approximation as a special case.

\subsection{Our results}

Despite broad use of the Hubbard-I approximation, its ability to account
for superconducting properties is untested. In this paper we address this
issue by comparing the Cooper pair propagator found with the help of
generalized Hubbard-I method against the same propagator calculated within
the MC framework (in this paper we refer to our calculations of the
two-particle Green's function as the {\it generalized} Hubbard-I
approximation to distinguish it from the classical Hubbard-I scheme
\cite{HubI}
designed for evaluation of the single-particle Green's function).

Such comparison has became possible only recently due to significant
progress of computational techniques
\cite{Ai,gauss}.
We use the MC data of Aimi and
Imada~\cite{Ai}, who analyze whether the Hubbard model is enough to capture
the phenomenon of high-Tc superconductivity. These authors use an advanced
numerical approach (pre-projected Gaussian-basis Monte Carlo). It evaluates
the equal-time two-particle propagator in $d_{x^2-y^2}$-channel
$P_d (r)$
without any prior assumptions about the structure of the ground state wave
function. The MC propagator on 10x10 square lattice does not exhibit
superconducting correlations up to $U=7t$. Instead, when $r$ grows,
$P_d (r)$
decays algebraically until finite size effects set in.

On our part, we generalize the classic Hubbard-I scheme
\cite{HubI} 
and derive the equation of motion for the Cooper pair propagator. We
determine
$P_d$
for the doped Hubbard model (\ref{Ham}). It is shown that at large
$U/t$
the results predicted by the Hubbard-I approach are qualitatively consistent
with numerical data obtained in Ref.~\cite{Ai}. In particular,
both approaches agree that the
$d_{x^2-y^2}$-wave
pairing correlations decay as
$c r^{-3}$
(``free electron'' behaviour). The Hubbard-I approximation underestimates
coefficient $c$ by a factor of order three. Our findings may be viewed as
an attempt to establish the limits of applicability of the Hubbard-I
scheme.

The paper is organized as follows. In Sect.~\ref{hubbard-i} we derive the
equation of motion for the two-particle propagator. Numerically calculated
Hubbard-I correlation function is presented and compared against the Monte
Carlo data in Sect.~\ref{numerics}. The results obtained are discussed in
Sect.~\ref{discussion}.

\section{Hubbard-I approximation} \label{hubbard-i}
\subsection{General idea}

The idea of the Hubbard-I approximation for Hamiltonian (\ref{Ham}) is
as follows. First, we introduce single-electron Matsubara Green's function
$G_\sigma(\mathbf{j-i}, \tau)
=
-\langle
	\hat{T}
	c^{\vphantom{\dagger}}_{\mathbf{j} \sigma}(\tau) 
	c^\dag_{\mathbf{i} \sigma} (0)
\rangle$,
where
$\langle ...\rangle$
means thermodynamic average,
$\hat{T}$
is the time-ordering operator, and $\tau$ is imaginary time. The equation of
motion for $G_\sigma(\mathbf{j-i},\tau)$ can be written as
\begin{eqnarray}
\label{Gree}
\nonumber
\left(
	-\frac{\partial}{\partial \tau}
	+
	\mu
\right)
G_\sigma(\mathbf{j-i}, \tau)
= 
\delta_{\mathbf{ji}} \delta( \tau)
+
UF_{\sigma\bar{\sigma}}(\mathbf{j-i}, \tau)
\\
-
t\sum_{\mathbf{a}}
	G_\sigma(\mathbf{j-i+a}, \tau),
\end{eqnarray}
where $\delta_{\mathbf{ij}}$ is the Kronecker symbol, $\delta(\tau)$ is the
delta-function, and $F_{\sigma\bar{\sigma}}(\mathbf{j-i},\tau)$ is the
two-particle Green's function of the form
\begin{equation}
\label{Gree2}
F_{\sigma\bar{\sigma}}(\mathbf{j-i}, \tau)
=
-\langle 
\hat{T}	c^{\vphantom{\dagger}}_{\mathbf{j} \sigma} (\tau)
	n^{\vphantom{\dagger}}_{\mathbf{j}\bar{\sigma}}(\tau)
	c^\dag_{\mathbf{i} \sigma}(0)
\rangle.
\end{equation}
Second, we write the equation of motion for function $F$, which includes
even more complicated Green's functions, which, in turn, require additional
equations of motion, etc. In the Hubbard-I approximation we avoid
proliferation of these Green's functions by making the following decoupling:
\begin{eqnarray}
\label{decoupl_dens}
\langle
	\hat{T} c_{\mathbf{j+a} \sigma}^{\vphantom{\dagger}} (\tau)
	n^{\vphantom{\dagger}}_{\mathbf{j}\bar{\sigma}}(\tau)
	c^\dag_{\mathbf{i} \sigma}(0)
\rangle
\rightarrow
\langle
	n_{\mathbf{j}\bar{\sigma}}
\rangle
\langle
	\hat{T} c_{\mathbf{j+a} \sigma}^{\vphantom{\dagger}} (\tau)
	c^\dag_{\mathbf{i} \sigma}(0)
\rangle,
\\
\label{decoupl_sc}
\langle \hat{T} c_{\mathbf{j} \sigma}^{\vphantom{\dagger}} (\tau)
		c_{\mathbf{j+a} \bar{\sigma}}^\dag (\tau)
		c_{\mathbf{j}\bar{\sigma}}^{\vphantom{\dagger}} (\tau)
		c^\dag_{\mathbf{i} \sigma}(0)
\rangle
\rightarrow
\langle
	c_{\mathbf{j}\bar{\sigma}}^{\vphantom{\dagger}}
	c_{\mathbf{j} \sigma}^{\vphantom{\dagger}}
\rangle
\\
\nonumber
\times
\langle
	\hat{T}
	c_{\mathbf{j+a} \bar{\sigma}}^\dag (\tau) 
	c^\dag_{\mathbf{i} \sigma}(0)
\rangle
= 0,
\\
\label{decoupl_spin}
\langle
	\hat{T} 
	c_{\mathbf{j} \sigma}^{\vphantom{\dagger}} (\tau)
	c_{\mathbf{j} \bar{\sigma}}^\dag (\tau)
	c_{\mathbf{j+a}\bar{\sigma}}^{\vphantom{\dagger}} (\tau)
	c^\dag_{\mathbf{i} \sigma}(0)
\rangle
\rightarrow
\langle
	c_{\mathbf{j} \sigma}^{\vphantom{\dagger}}
	c_{\mathbf{j} \bar{\sigma}}^\dag
\rangle
\\
\nonumber
\times
\langle
	\hat{T}
	c_{\mathbf{j+a}\bar{\sigma}}^{\vphantom{\dagger}} (\tau) 
	c^\dag_{\mathbf{i} \sigma}(0)
\rangle
= 0. 
\end{eqnarray}
To understand the nature of this approximation let us express the electron
density as a sum of its average value and fluctuations around it:
\begin{eqnarray}
n_{\mathbf{j}\bar{\sigma}}
=
\langle
	n_{\mathbf{j}\bar{\sigma}}
\rangle
+
\delta n_{\mathbf{j}\bar{\sigma}}, {\rm \ where}
\\
\delta n_{\mathbf{j}\bar{\sigma}}
=
n_{\mathbf{j}\bar{\sigma}}
-
\langle
	n_{\mathbf{j}\bar{\sigma}}
\rangle.
\end{eqnarray}
Two other boson-like quantities,
$
c_{\mathbf{j}\bar{\sigma}}^{\vphantom{\dagger}}
c_{\mathbf{j} \sigma}^{\vphantom{\dagger}}
$
and
$
c_{\mathbf{j} \sigma}^{\vphantom{\dagger}}
c_{\mathbf{j} \bar{\sigma}}^\dag
$,
can be represented in the same manner. The above decouplings correspond to
the assumption that the propagation of a single electron is unaffected by
the fluctuations of these quantities around their average values. This
assumption could be proven by either showing that {\it (i)} the fluctuation
terms were small, or that {\it (ii)} there were no correlation between the
single-electron motion and the fluctuations. Unfortunately, neither {\it
(i)} nor {\it (ii)} are rigorously established. Because of this the
Hubbard-I approximation is uncontrollable approach whose accuracy is
unknown.

Using the decouplings 
(\ref{decoupl_dens}-\ref{decoupl_spin})
we derive the equation for $F$ in the form
\begin{eqnarray}
\label{Gree22} 
\nonumber 
\left(
	-\frac{\partial}{\partial \tau}
	+
	\mu
	-
	U
\right)
F_{\sigma\bar{\sigma}}(\mathbf{j-i},\tau)
\\
\nonumber
=
\langle
	n_{\mathbf{j}\bar{\sigma}}
\rangle
\left[
	\delta_{\mathbf{ji}}
	\delta(\tau)
	-
	t\sum_{\mathbf{a}}
		G_\sigma(\mathbf{j-i+a},\tau)
\right],
\end{eqnarray}
This equation, together with
(\ref{Gree}),
constitutes a closed system, sufficient for calculation of functions $G$ and
$F$. This is how the single-electron propagator is calculated within the
framework of the Hubbard-I approximation.

\subsection{Hubbard-I approximation for the two-particle propagator}

Following Ref.~\cite{Ai}, we calculate here the equal-time
two-electron correlation functions of the form
\begin{equation}\label{PP}
P_d(\mathbf{r})
=
\frac{1}{2N}
\sum_{\mathbf{i}=1}^N
\langle
	\Delta^\dag_d(\mathbf{i})
	\Delta^{\vphantom{\dagger}}_d(\mathbf{i}+\mathbf{r})
	+
	\Delta_d(\mathbf{i})^{\vphantom{\dagger}}
	\Delta^\dag_d(\mathbf{i}+\mathbf{r})
\rangle,
\end{equation}
where $N$ is the number of sites on the square lattice,
\begin{eqnarray}
\label{Delta}
\Delta_d(\mathbf{i})
=
\frac{1}{\sqrt{2}}
\sum_\mathbf{r}
f_d(\mathbf{r})
\left( 
	\Delta_{\mathbf{i i+r}} 
	+ 
	\Delta_{\mathbf{i+r i}}
\right),
\\
\Delta_{\mathbf{ij}}= c^\dag_{{\mathbf{i}\sigma}}
c^\dag_{\mathbf{j}\bar{\sigma}}. 
\end{eqnarray}
The form-factor
\begin{equation}
\label{f_d}
f_d(\mathbf{r})=
\delta_{r_y,0}\left(\delta_{r_x,1}+\delta_{r_x,-1}\right)
-\delta_{r_x,0}\left(\delta_{r_y,1}+\delta_{r_y,-1}\right)
\end{equation}
corresponds to the
$d_{x^2-y^2}$-wave
order parameter. 

To calculate 
$P_d ({\bf r})$
let us first define the propagator 
\begin{eqnarray}
P_{\mathbf{ijlm}}(\tau) 
=
\langle
	\hat{T}
	\Delta_{\mathbf{ij}}^{\vphantom{\dagger}} (\tau)
	\Delta^\dag_{\mathbf{lm}}(0)
\rangle.
\end{eqnarray}
The equal-time two-electron correlation function
$P_d(\mathbf{r})$
is related to
$P_{\mathbf{ijlm}}(\tau)$
as
\begin{eqnarray}
\label{P-P}
P_d(\mathbf{r})
=
\frac{1}{4N}
\sum_{\mathbf{ibb'}}
	f_d(\mathbf{b})
	f_d(\mathbf{b'})
	\Big[
	P_{\mathbf{i, i+b, i+r, i+b'+r}}(\tau)
\\
\nonumber
	+
	P_{\mathbf{i+b, i, i+r, i+b'+r}}(\tau)
	+
	P_{\mathbf{i, i+b, i+b'+r, i+r}}(\tau)
\\
\nonumber 
	+ 
	P_{\mathbf{i+b, i, i+b'+r, i+r}}(\tau)
\Big]\Big|_{\tau \rightarrow -0}
+
\left(
	\tau \rightarrow +0,\ {\bf r} \rightarrow -{\bf r}
\right).
\end{eqnarray}
We write down equation of motion for the propagator
$P_{\mathbf{ijlm}}$
using Hamiltonian (\ref{Ham}) and (\ref{Ham1})
\begin{eqnarray}
\label{P1}
\nonumber
\left(
	\frac{\partial}{\partial \tau}
	+
	2\mu
\right)
P_{\mathbf{ijlm}} 
=
-t\sum_\mathbf{a}
	\left(
		P_{\mathbf{i+a jlm}}
		+
		P_{\mathbf{ij+a lm}}
	\right)
\\
	+
	\delta(\tau)
	\left(
		\langle 
			c^\dag_{\mathbf{i}\sigma} 
			c_{\mathbf{l}\sigma}^{\vphantom{\dagger}}
		\rangle
		\delta_{\mathbf{jm}}
		-
		\langle
			c_{\mathbf{m}\bar{\sigma}}^{\vphantom{\dagger}}
			c_{\mathbf{j}\bar{\sigma}}^\dag
		\rangle
		\delta_{\mathbf{il}}
	\right)
	+
	U{\cal P}_{\mathbf{ijlm}},
\end{eqnarray}
where the three-particle propagator is defined as
\begin{equation}
\label{calP}
{\cal P}_{\mathbf{ijlm}}(\tau)
=
\langle
	\hat{T}
	\left[
		\Delta_{\mathbf{ij}}^{\vphantom{\dagger}} (\tau)
		n_{\mathbf{j}\sigma}^{\vphantom{\dagger}} (\tau)
		+
		n_{\mathbf{i}\bar{\sigma}}^{\vphantom{\dagger}} (\tau)
		\Delta_{\mathbf{ij}}^{\vphantom{\dagger}} (\tau)
	\right]
	\Delta^\dag_{\mathbf{lm}}(0)
\rangle.
\end{equation}
The double occupancy is very unlikely: 
$P_{\mathbf{ijlm}}=o(t/U)\approx 0$
if
$\mathbf{i}=\mathbf{j}$
or
$\mathbf{l}=\mathbf{m}$.
Assuming the absence of the magnetic order, which means, in particular,
$\langle
	c_{\mathbf{i}\sigma}^{\vphantom{\dagger}} 
	c^\dag_{\mathbf{j}\sigma}\rangle= \langle
	c_{\mathbf{i}\bar{\sigma}}^{\vphantom{\dagger}} 
	c^\dag_{\mathbf{j}\bar{\sigma}}
\rangle$,
we transform (\ref{P1}) in the Hubbard-I approximation to
\begin{eqnarray}
\label{P11}
\nonumber
\left(
	\!\frac{\partial}{\partial \tau}\!
	+
	\!2\mu\!
\right)
P_{\mathbf{ijlm}}\! 
=\!
\!-\!
t\sum_\mathbf{a}
	\left(
		P_{\mathbf{i+a jlm}}\!
		+\! 
		P_{\mathbf{ij-a lm}}
	\right)
	\left(
		1\!-\!\delta_{\mathbf{i+a j}}
	\right)\!
\\
\!\!\!\!
+ \delta(\tau)\!
\left(
	\langle 
		c^\dag_{\mathbf{i}\sigma} 
		c_{\mathbf{l}\sigma}^{\vphantom{\dagger}}
	\rangle
	\delta_{\mathbf{jm}}
	-
	\langle
		c_{\mathbf{m} {\sigma}}^{\vphantom{\dagger}}
		c_{\mathbf{j} {\sigma}}^\dag
	\rangle
	\delta_{\mathbf{il}}
\right)
+
U{\cal P}_{\mathbf{ijlm}},\quad
\end{eqnarray}
which is valid if
$\mathbf{i}\neq \mathbf{j}$ and $\mathbf{l}\neq \mathbf{m}$.

Now we need to derive an equation for 
${\cal P}$.
The required calculations are onerous, but straightforward. They are
presented in
\ref{derivation_of_PPP}.
Here we quote the final result:
\begin{eqnarray}
\label{PPP}
U{\cal P}_{\mathbf{ijlm}}\!
\approx
t\!\langle n_{\mathbf{i}\sigma}\rangle
\sum_{\mathbf{a}}\!
	\left(
		P_{\mathbf{i+a jlm}}\!
		+\! 
		P_{\mathbf{ij-a lm}}
	\right)
	\left(
		1\!-\!\delta_{\mathbf{i+a j}}
	\right)\!
\\
\nonumber
-
\delta(\tau)
\langle n_{\mathbf{i}\sigma}\rangle
\left(
	\langle 
		c^\dag_{\mathbf{i}\sigma} 
		c_{\mathbf{l}\sigma}^{\vphantom{\dagger}}
	\rangle
	\delta_{\mathbf{jm}}
	-
	\langle
		c_{\mathbf{m} {\sigma}}^{\vphantom{\dagger}}
		c_{\mathbf{j} {\sigma}}^\dag
	\rangle
	\delta_{\mathbf{il}}
\right).
\end{eqnarray}
This relation is derived under the assumption that the magnetic order is
absent:
$\langle n_{{\bf i} \sigma} \rangle
=
\langle n_{{\bf i} \bar{\sigma}} \rangle$.

Substituting (\ref{PPP}) in (\ref{P11}), we obtain
the equation of motion for the propagator
$P_{\mathbf{ijlm}}$
in the Hubbard-I approximation valid when
$\mathbf{i}\neq \mathbf{j}$
and
$\mathbf{l}\neq \mathbf{m}$
\begin{eqnarray}
\label{PHI}
\nonumber
\left(
	\frac{\partial}{\partial \tau}
	+
	2\mu
\right)P_{\mathbf{ijlm}}
=
\\
\nonumber
-\tilde{t}
\sum_{\mathbf{a}}
	\left[
		P_{\mathbf{i+a jlm}}
		\left(1\!-\!\delta_{\mathbf{ji+a}}\right)\!
		+\!
		P_{\mathbf{ij+a lm}}
		\left(1\!-\!\delta_{\mathbf{ij+a}}\right)
	\right]
\\
+
\delta(\tau)
\left(1-\langle
n_{\mathbf{i}\sigma}\rangle\right)
\left(
	\langle 
		c^\dag_{\mathbf{i}\sigma} 
		c_{\mathbf{l}\sigma}^{\vphantom{\dagger}}
	\rangle
	\delta_{\mathbf{jm}}
	-
	\langle
		c_{\mathbf{m} {\sigma}}^{\vphantom{\dagger}}
		c_{\mathbf{j} {\sigma}}^\dag
	\rangle
	\delta_{\mathbf{il}}
\right).
\end{eqnarray}
Here
$\tilde{t}=t\left(1-\langle n_{\mathbf{i}\sigma}\rangle\right)$
if the system is uniform.

As we wrote above,
$P_{\mathbf{iilm}}\sim o(t/U)\approx 0$
within our accuracy. Thus, jumps from site
$\mathbf{j}=\mathbf{i+a}$
to site
$\mathbf{i}$
must be excluded. The same is true for jumps from
$\mathbf{i}=\mathbf{j+a}$
to
$\mathbf{j}$.
To take this fact into account and to generalize the last equation for
the case
$\mathbf{i}=\mathbf{j}$,
we add the term
$t(1-\langle n_{{\bf i}\sigma} \rangle)
\sum_{\mathbf{a}}
\left(
	\delta_{\mathbf{i+a j}}
	P_{\mathbf{i+a jlm}}
	+
	\delta_{\mathbf{ij+a}}P_{\mathbf{ij+a lm}}
\right)$
to the right-hand side of (\ref{PHI}). Therefore:
\begin{eqnarray}
\label{PHI1}
\nonumber
\left(
	\frac{\partial}{\partial \tau}
	+
	2\mu
\right)
P_{\mathbf{ijlm}} =
\\
\nonumber
-\tilde{t}
\sum_{\mathbf{a}}
\left[
	P_{\mathbf{i+a jlm}}
	\left(
		1\!
		-\!
		\delta_{\mathbf{ji+a}}\!
		-\!
		\delta_{\mathbf{ij}}
	\right)\!
	+\!
	P_{\mathbf{ij+a lm}}
	\left(
		1\!
		-\!
		\delta_{\mathbf{ij+a}}\!
		-\!
		\delta_{\mathbf{ij}}
	\right)
\right]
\\
+
\delta(\tau)
\left(1-\langle n_{\mathbf{i}\sigma}\rangle\right)
\left(
	\langle 
		c^\dag_{\mathbf{i}\sigma} 
		c_{\mathbf{l}\sigma}^{\vphantom{\dagger}}
	\rangle
	\delta_{\mathbf{jm}}
	-
	\langle
		c_{\mathbf{m} {\sigma}}^{\vphantom{\dagger}}
		c_{\mathbf{j} {\sigma}}^\dag
	\rangle
	\delta_{\mathbf{il}}
\right).
\end{eqnarray}
In such equation the quantity $P_{\mathbf{iilm}}$ is
decoupled from $P_{\mathbf{ijlm}}$, $\mathbf{i} \ne \mathbf{j}$, which is
equivalent to the condition $P_{\mathbf{iilm}}=0$. 

Expression (\ref{PHI1}) corresponds to propagation of two interacting
particles: the terms with the Kronecker symbols may be regarded as an
effective repulsive interaction between electrons on neighboring sites with
the coupling constant of the order of
$t(1-\langle n_{{\bf i}\sigma} \rangle )$.
Thus, we reduce the strong-coupling many-body problem to the
intermediate-coupling two-body one.

\begin{figure}[btp]
\begin{center}
\includegraphics[width=8.5 cm]{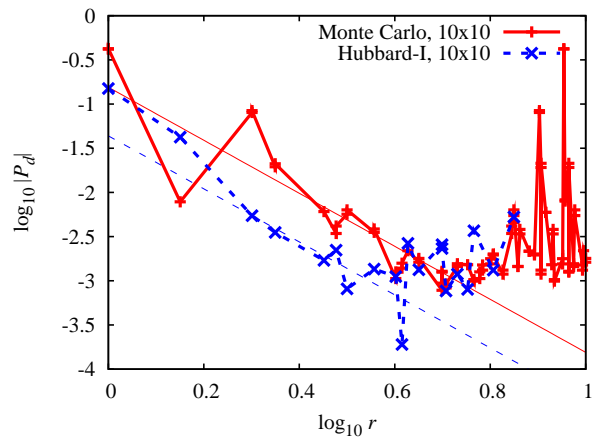}
\\
\end{center}
\caption {(Colour online) The two-particle propagator
$P_d(r)$,
equation~(\ref{PP}).
Solid (red) line with right crosses is the Monte Carlo data for the doping
level of 0.18 and
$U=6$
collected on 10x10 lattice,
Ref.~\cite{Ai}.
Dashed (blue) line with skew crosses is the Hubbard-I result for the doping
level of 0.17 on 10x10 lattice. The two straight lines correspond to
$c/r^3$
behaviour.
}
\label{10x10}
\end{figure}

\begin{figure}[btp]
\begin{center}
\includegraphics[width=8.5 cm]{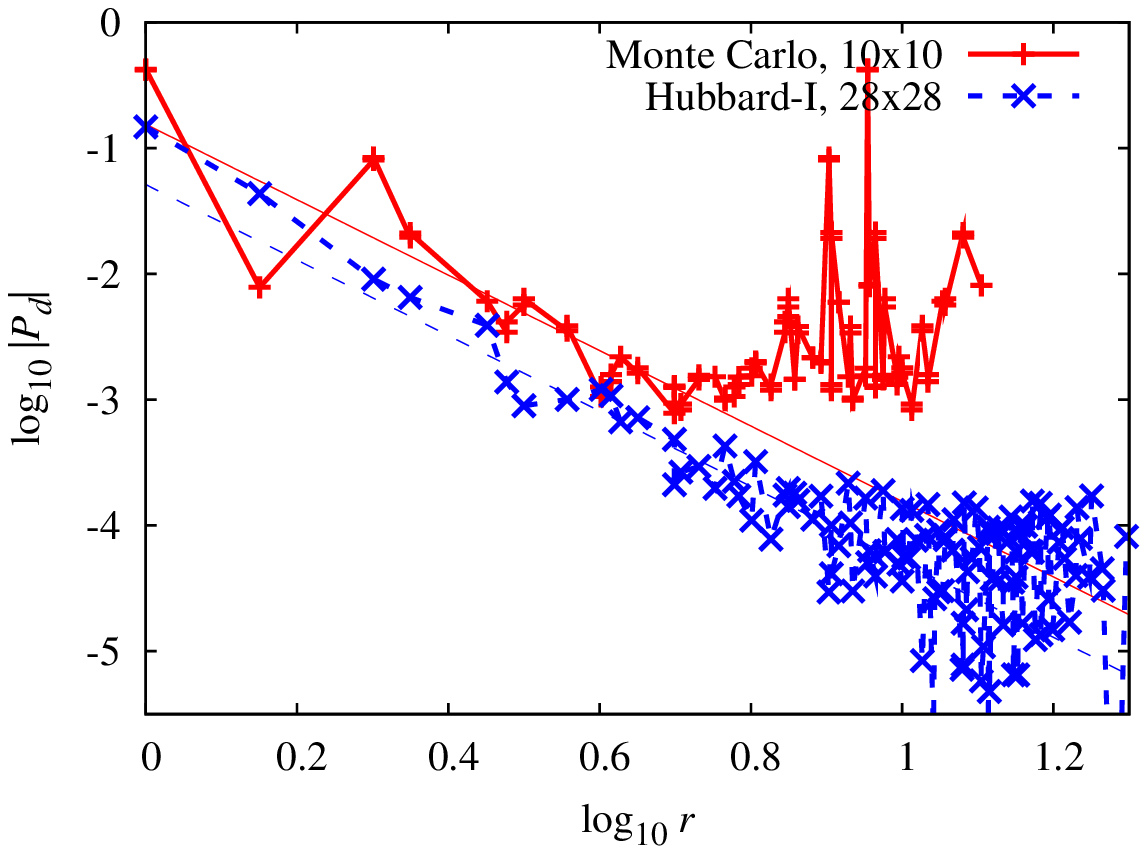}
\\
\end{center}
\caption {(Colour online) The two-particle propagator
$P_d(r)$,
equation~(\ref{PP}).
Solid (red) line with right crosses is the Monte Carlo data for the doping
level of 0.18 and
$U=6$
collected on 10x10 lattice,
Ref.~\cite{Ai}.
Dashed (blue) line with skew crosses is the Hubbard-I result for the doping
level of 0.18 on 28x28 lattice. Two straight lines correspond to
$c/r^3$
behaviour.}
\label{fit}
\end{figure}

\begin{figure}[btp]
\begin{center}
\includegraphics[width=8.5 cm]{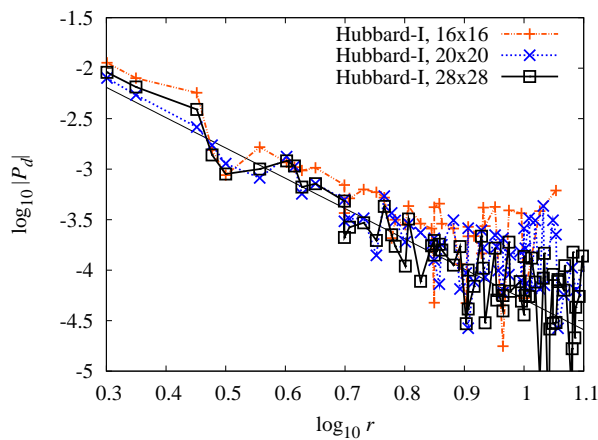}
\\
\end{center}
\caption {(Colour online) 
Finite-size scaling for the Hubbard-I propagator: the data for the lattices
of three different sizes (16x16, 20x20, and 28x28) and the doping levels
close to 0.18 are plotted. While the curve for the smallest lattice (brown
curve with right crosses) deviates substantially from the remaining two,
the latter are quite close to each other for sufficiently small separations
($\log_{10} r < 0.8$).}
\label{scaling}
\end{figure}


\section{Numerical results} \label{numerics}

\subsection{Hubbard-I vs. MC}

In this section we present the results of our numerical calculation of the
equal-time two-electron correlation function 
(\ref{PP})
at a given doping level. The actual calculations are based on
Eq.~(\ref{Solu1})
derived in 
\ref{fourier}
and~\ref{comput} 
Doping is equal to
$1 - n$,
where the average number of electrons per site is defined as
$n=\sum_{\sigma} \langle n_{\mathbf{i}\sigma}\rangle$.
The average
$\langle n_{\mathbf{i}\sigma}\rangle$
within the Hubbard-I approximation is
\begin{eqnarray}
\langle n_{\mathbf{i}\sigma}\rangle
=
( 1 - \langle n_{\mathbf{i}\sigma}\rangle ) n_0,
\\
n_0 (\mu)
=
\frac{
	1
     }
     {
	N
     }
\sum_{\bf k} n_{\rm F} (\tilde \varepsilon_{\bf k} - \mu).
\end{eqnarray}
From this equation the expression for $n$ is found:
\begin{eqnarray}
\label{n_phys}
n = \frac{ 2 n_0 (\mu) }{ 1 + n_0 (\mu) }.
\end{eqnarray}
Therefore, the doping level can be adjusted by changing $\mu$.

For calculation at zero temperature it is convenient to use the
substitution in (\ref{Solu1}):
\begin{eqnarray}
N_{\rm B} (\varepsilon)
=
\frac{1}{{\rm sh}\, ( \varepsilon / T)} 
- 
n_{\rm F} (\varepsilon).
\end{eqnarray} 
At $T=0$ the Fermi distribution function is zero (unity) for
$\varepsilon>0$
($\varepsilon < 0$). 
Contribution form 
$1/{\rm sh}\, ( \varepsilon / T)$
vanishes at small temperature.

Numerical results are shown in
figure~\ref{10x10} and figure~\ref{fit},
where the propagator is plotted versus separation
$r=|\mathbf{r}|$
on the log-log graph. The Hubbard-I propagator 
$P_{HI}$
is shown as the dashed (blue) line. For comparison, the same propagator
computed by
T.~Aimi and M.~Imada~\cite{Ai}
using the assumption-free quantum Monte Carlo procedure,
$P_{MC}$,
is shown by solid (red) line. The subscript `HI' stands for `Hubbard-I',
the subscript `MC' -- for `Monte Carlo'.

In figure~\ref{10x10} we present the Hubbard-I data calculated on a 10x10
lattice for the doping level of 0.17. The same lattice size is used in the
MC simulation of
Ref.~\cite{Ai}.
The MC data is collected for the doping level of 0.18. The small
discrepancy in the densities is unavoidable in our situation. Indeed, due
to the finite size of the system, the number of particles cannot be changed
continuously. In addition, in case of Hubbard-I scheme, the number of
physical electrons is not an independent quantity, but rather has to be
calculated according to
formula~(\ref{n_phys}).
Given these two circumstances, it is impossible in general to match the
MC and Hubbard-I densities exactly. However, such a small discrepancy
(less than 6\%) is of little importance for our goal.

The solid straight line fits the MC propagator in the window 
$0.32 < \log_{10} r < 0.78$.
Such fit corresponds to 
$c/r^3$
decay, as discussed in
Ref.~\cite{Ai}.
Propagator
$P_{HI}$
is also fited by a straight line (blue dashed straight line) in the window 
$0.3 < \log_{10} r < 0.6$. 

To extract the asymptotic behaviour of 
$P_{HI} (r)$
more reliably it is better to look at the propagator on a bigger lattice.
In
figure~\ref{scaling}
it is shown how the propagator converges for large lattices. We observe
that the curves for the lattices with
$M \geq 20$
almost coincide for moderate $r$. Thus, such systems can be used to
find the asymptotic. To that end, let us examine
figure~\ref{fit}.
It presents the Hubbard-I data for 28x28 lattice is shown together with the
MC data.  Analysis of
figure~\ref{fit}
reveals that the difference between assumption-free MC and much more simple
Hubbard-I calculations is not large, moreover, qualitative appearances of
the curves are similar. In particular, both curves decay as a power-law
$c/r^\nu$.
In~\cite{Ai}
it is established that 
$\nu\approx 3$,
which is consistent with the free-electron behaviour. To determine this
exponent for the Hubbard-I correlation function we use two-parameter
least-square fitting procedure for the Hubbard-I data in the range
$0.3 < \log_{10} r < 0.9$.
The fitting returns the following values:
\begin{eqnarray}
\log_{10} c_{\rm FIT} = -1.2 \pm 0.1,\ \nu_{\rm FIT} = 3.1 \pm 0.2.
\end{eqnarray} 
The value of $\nu_{\rm FIT}$ is consistent with the MC result. Below we
will always assume that the Hubbard-I correlation function decays with 
$\nu=3$.
Deviations from the free-electron behaviour at large $r$ are a
manifestation of the finite-size effects. 

The fact that the MC propagator decays as the free-electron propagator is a
surprising revelation of 
Ref.~\cite{Ai}.
The ability of the Hubbard-I approximation to capture the
$1/r^3$
law is a rather simple consequence of the equation of motion,
(\ref{PHI1}).
Indeed, in the latter equation, if $r$ is large, the interaction terms (the
terms with the Kronecker symbols) contribute to the $s$-wave channel only;
as for the $d$-wave propagator, its equation of motion coincides, up to a
normalization factor, with that for free fermions.

In
figure~\ref{fit}
the obtained values of the propagators are ``noisy''. These oscillations
arise, because, in general, the propagators depend not only on the distance
$r$, but on the direction of the vector
$\mathbf{r}$
as well. Despite the ``noise'', one can easily observe that the Hubbard-I
propagator
$P_{HI}$
is smaller than the Monte Carlo propagator $P_{MC}$. 

To examine
$P_{HI}$
and
$P_{MC}$
in a more rigorous manner let us quantify the
$1/r^3$
behaviour of the propagators. For this aim we fit both functions in
figure~\ref{fit} by straight lines in the window 
$0.32 < \log_{10} r < 0.78$.
The fit lines are defined by the equations:
\begin{eqnarray}
\log_{10} P_{MC} 
=
-0.81-3 \log_{10} r,
\\
\log_{10} P_{HI}
=
-1.29-3 \log_{10} r,
\end{eqnarray} 
which correspond to
$P_d(r)\approx c/r^3$
with
$c_{MC} \approx 0.15$,
$c_{HI} \approx 0.051$,
and the ratio
$c_{MC}/c_{HI} \approx 2.9$. 
Thus, the Hubbard-I propagator's asymptotic coincides with the asymptotic
of the propagator calculated with the help of the assumption-free Monte
Carlo method within an order of magnitude, but, of course, not exactly.

\begin{figure}[btp]
\begin{center}
\includegraphics[width=8.5 cm]{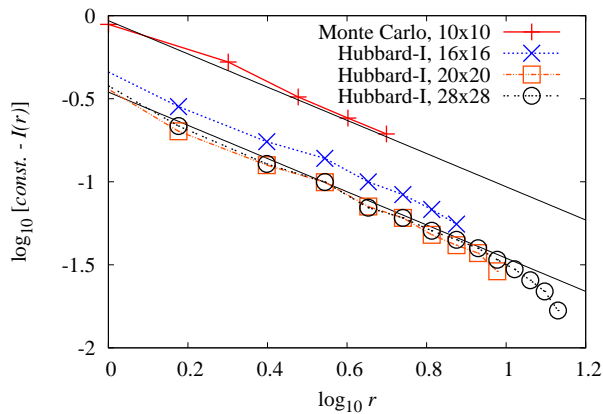}\\
\end{center}
\caption {(Colour online) Function
$[\alpha - I(r)]$,
equation~(\ref{Ir});
solid (red) line with right crosses is the Monte Carlo data. Three other
lines are the Hubbard-I results for different lattice sizes. The parameters
are the same as in
figure~\ref{fit}.
Two straight lines fit the propagators at moderate $r$ according to
equation~(\ref{fitI}).
The values of $\alpha$'s are as follows:
$\alpha_{MC} = 4.1$,
$\alpha_{HI}^{16} = 1.45$,
$\alpha_{HI}^{20} \approx \alpha_{HI}^{28} = 1.40$.
 }
\label{integ}
\end{figure}

We can further compare the Hubbard-I and Monte Carlo data by eliminating
``noise'' in the propagators with the help of the following procedure. 
Define the sum
\begin{eqnarray}\label{Ir} I(r) = \sum_{\sqrt{x^2+y^2} < r}
P_d(x,y). 
\end{eqnarray}
If we assume that $P_d(x,y) \sim c(\phi) r^{-3}$,
where $\phi$ is the polar angle, then $I(r)$ can be approximated as
\begin{equation}
\label{inte}
 I(r)
 \approx
 \int_{x^2+y^2 < r^2} dx dy P(x,y)
\approx \alpha - {\beta \over r},
\end{equation}
where
\begin{equation}
\label{bc}
\beta=\int_0^{2\pi}c(\phi)d\phi=2\pi \bar{c},
\end{equation}
and the constant $\alpha$ can be estimated as
$\alpha \approx I(r_{\rm max})$,
with
$r_{max}=M/2$.
Function 
$I(r)$,
by its definition, is independent of vector 
${\bf r}$'s
direction. Consequently, it is smooth function, free from irregular
oscillations present in
figure~\ref{fit}.
This makes the fitting procedure more robust.

Dependence
$\log_{10} (\alpha-I)$
vs.
$\log_{10} r$
is plotted in
figure~\ref{integ}
for the same values of parameters as the data in
figure~\ref{fit}.
Similarly to the data presented in
figure~\ref{scaling},
the 16x16 curve lies somewhat away from the curves corresponding to the
larger lattices. The curves for 20x20 and 28x28 lattices coincide with each
other almost everywhere, suggesting that for systems of such sizes the
finite-size effects are important for large $r$ only.

To determine the asymptotic parameters the curves in
figure~\ref{integ}
are fitted by the straight lines:
\begin{eqnarray}
\log_{10} (\alpha-I) = \log_{10} 2\pi \bar{c} - \log_{10} r.
\label{fitI}
\end{eqnarray}
The values of the fitting constants are:
$\bar{c}_{MC} \approx 0.15$,
$\bar{c}_{HI} \approx 0.055$.
As expected, they are close to
$c_{MC}$
and
$c_{HI}$.

\begin{figure}[btp]
\begin{center}
\includegraphics[width=8.5 cm]{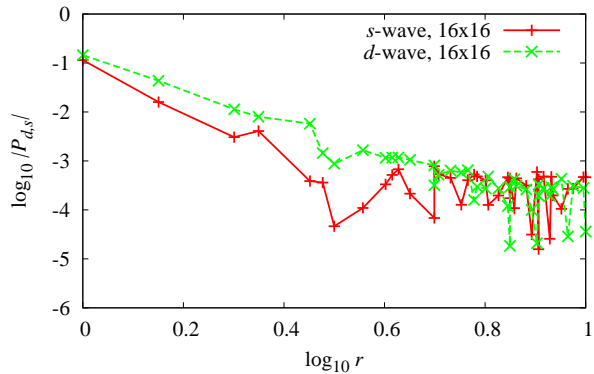}\\
\end{center}
\caption {(Colour online) Dependence of
$d$-wave and $s$-wave propagators on $r$; $P_d(r)$ -- dashed (green) line and
$P_s(r)$ -- solid (red) line. Both propagators are calculated with the help
of Hubbard-I approximation for the doping level of 0.19 on 16x16 lattice.}
\label{fig4}
\end{figure}

\subsection{Application}
The numerical procedure in the case of the Hubbard-I approximation is much
simpler than the quantum Monte Carlo computations. One can use it to derive
different properties of the Hamiltonian~(\ref{Ham1}). As an example, we
apply the Hubbard-I approximation (\ref{FinP}) to calculation of the
equal-time two-electron correlation function in the $s$-channel:
\begin{equation}
\label{Ps}
P_s(\mathbf{r})
=
\frac{1}{2N}
\sum_{\mathbf{i}=1}^N
\langle
	\Delta^\dag_s(\mathbf{i})
	\Delta_s(\mathbf{i}+\mathbf{r})
	+
	\Delta_s(\mathbf{i})
	\Delta^\dag_s(\mathbf{i}+\mathbf{r})
\rangle,
\end{equation}
where
\begin{equation}
\label{DeltaS}
\Delta_s(\mathbf{i})
=
\frac{1}{\sqrt{2}}
\sum_\mathbf{r}
f_s(\mathbf{r})
\left( 
	\Delta_{\mathbf{i i+r}} 
	+ 
	\Delta_{\mathbf{i+r i}}
\right),
\end{equation}
and $f_s$ is the $s$-wave form-factor
\begin{equation}
\label{f_s}
f_s(\mathbf{r})
=
\delta_{r_y0}
\left(
	\delta_{r_x1}
	+
	\delta_{r_x-1}
\right)
+
\delta_{r_x0}
\left(
	\delta_{r_y1}	
	+
	\delta_{r_y-1}
\right).
\end{equation}
The result is shown in
figure~\ref{fig4}
by solid (red) line. Function
$P_s(r)$
decays quickly for small $r$. At larger $r$ it oscillates around
finite-value plateau. As could be expected, the $s$-wave superconducting 
correlations are weaker and decay faster than the $d$-wave ones [compare
solid (red) and dashed (green) lines in
figure~\ref{fig4}].
This is because the electrons in $s$ channel experience strong repulsion
suppressing the superconducting correlations.

Note also that the Hubbard-I correlation functions show no sign of the
superconducting instability in both $s$ and $d$ channels: both correlation
functions decay at large $r$, while in a superconducting phase the order
parameter correlation function must saturate at large distances.

\section{Discussion}
\label{discussion}

We calculated two-electron equal-time $d$-wave propagator using the
Hubbard-I approximation and compared the results with quantum Monte Carlo
computations~\cite{Ai}. The results of both approaches are in good
qualitative agreement. Specifically, both methods predict that the correlation
function decays as $1/r^3$ at $r$ exceeding several lattice constants. This
allows us to assess the reliability of the Hubbard-I approximation.

Technically speaking, the
$1/r^3$
decay of the Hubbard-I propagator is a consequence of the fact that the
interaction in the equation of motion, (\ref{PHI1}), is important in
the $s$-channel only, whereas in the $d$-channel its effect decreases when
$r$ increases.

The Hubbard-I approximation underestimates the value of the propagator
residue by a factor of order three. Such discrepancy is rather reasonable for
an uncontrollable approach. 

Our results for the two-particle correlation functions suggest that
the Fermi liquid is a poor approximation for the Hubbard model near
half-filling. We show that in $d$-wave channel the correlation function
decays as $r^{-3}$, which is consistent with ``free fermions". However,
fast decay of the $s$-wave propagator (see figure~\ref{fig4}) does not agree
with the Fermi liquid picture. One, however, must remember that, unlike the
$d$-wave propagator, the Hubbard-I $s$-wave correlation function is not
compared against a controllable method, thus, its accuracy is unknown.

Interpreting our calculation one should keep in mind the following
restriction. Since the MC data for large $r$ is not available, we cannot
say how Hubbard-I approximation fairs for $r$ exceeding 5-6 lattice
constants. The MC data of 
Ref.~\cite{Ai}
cannot be used to rule out the possibility that the Hubbard model exhibits
superconductivity with the correlation length larger than 5-6 lattice
constants. Our method in its present form is too crude to capture the
superconducting correlations.

To conclude, generalizing the Hubbard-I approximation we derive the
equation of motion for the two-particle propagator. For the square lattice
of finite size this equation is solved numerically. It is demonstrated that
the Hubbard-I propagator is consistent with the MC data up to a numerical
coefficient of order 3. Thus, Hubbard-I method can be used to approximate
the two-particle propagator of the Hubbard model, at least at a qualitative
level at moderate separations.

\ack

The authors would like to thank Prof. Imada who kindly provided the Monte
Carlo data shown in
figure~\ref{10x10}
and
figure~\ref{fit}.
Discussions with M.Yu.~Kagan are gratefully acknowledged.
The support of RFBR (grants 09-02-00248, 09-02-92114, 08-02-00212) is
acknowledged.

\appendix

\section{Derivation of Eq.~(\ref{PPP})}
\label{derivation_of_PPP}
\setcounter{section}{1}

Here we obtain 
Eq.~(\ref{PPP})
assuming that
$\mathbf{i}\neq \mathbf{j}$
and
$\mathbf{l}\neq \mathbf{m}$. 
The equation of motion for
${\cal P_{\mathbf{ijlm}}}$ is
\begin{eqnarray}
\label{PPP1}
\dot{{\cal P}}_{\mathbf{ijlm}}
=
\left\langle
	\hat{T} C (\tau)
	      \Delta^\dag_{\mathbf{lm}}(0)
\right\rangle
+
\delta(\tau)
\left\langle 
	\left[
		\left(
			n_{\mathbf{i}\bar{\sigma}}
			+
			n_{\mathbf{j}\sigma}
		\right)
		\Delta_{\mathbf{ij}}^{\vphantom{\dagger}},
		\Delta^\dag_{\mathbf{lm}}
	\right]
\right\rangle,
\end{eqnarray}
where we introduced the notation:
\begin{eqnarray} 
C
=
\left[
	H,\left(
		n_{\mathbf{i}\bar{\sigma}} 
		+
		n_{\mathbf{j}\sigma} 
	  \right)
	  \Delta_{\mathbf{ij}}^{\vphantom{\dagger}}  
\right]
\\
\nonumber
=
\left[
	\left(
		H_{\rm kin}
		+
		H_{\rm int}
	\right),
	(n_{\mathbf{i}\bar{\sigma}}+n_{\mathbf{j}\sigma})
	\Delta_{\mathbf{ij}}
\right]
-
2\mu(n_{\mathbf{i}\bar{\sigma}}+n_{\mathbf{j}\sigma})
\Delta_{\mathbf{ij}}.
\end{eqnarray}
Neglecting ``double occupancy'' terms in the commutator with $H_{\rm kin}$,
we get
\begin{eqnarray}
\label{Hk}
\nonumber
\left[
	H_{\rm kin},
	(n_{\mathbf{i}\bar{\sigma}}+n_{\mathbf{j}\sigma})
	\Delta_{\mathbf{ij}}
\right]
\approx
-t\sum_\mathbf{a}
	n_{\mathbf{i}\bar{\sigma}}
	c^\dag_{\mathbf{i+a}\sigma}
	c^\dag_{\mathbf{j}\bar{\sigma}}
	\left(1-\delta_{\mathbf{ji+a}}
	\right)
\\
	+
	n_{\mathbf{j}\sigma}
	c^\dag_{\mathbf{i}\sigma}
	c^\dag_{\mathbf{j+a}\bar{\sigma}}
	\left(1-\delta_{\mathbf{j+a i}}\right)
	+
	c^\dag_{\mathbf{i+a}\bar{\sigma}}
	c_{\mathbf{i}\bar{\sigma}}
	c^\dag_{\mathbf{i}\sigma}
	c^\dag_{\mathbf{j}\bar{\sigma}}
	\left(1-\delta_{\mathbf{ji+a}}\right)
\,\,\,\,\\
\nonumber
	+
	c^\dag_{\mathbf{j+a}\sigma}
	c_{\mathbf{j}\sigma}
	c^\dag_{\mathbf{i}\sigma}
	c^\dag_{\mathbf{j}\bar{\sigma}}
	\left(1-\delta_{\mathbf{j+a i}}\right).
\qquad\qquad\qquad\qquad\qquad
\end{eqnarray}
The factors of the
type $1-\delta_{\mathbf{ij}}$ appear for the following reason. Consider:
\begin{eqnarray}
n_{\mathbf{j}\sigma} c^\dag_{\mathbf{i}\sigma}
c^\dag_{\mathbf{j+a}\bar{\sigma}}
&=&
n_{\mathbf{j}\sigma}c^\dag_{\mathbf{i}\sigma}
c^\dag_{\mathbf{j+a}\bar{\sigma}}
\left(
	1-\delta_{\mathbf{ij+a}}
\right)
\\
\nonumber
&-&
n_{\mathbf{j}\sigma}
c^\dag_{\mathbf{i}\sigma}
c^\dag_{\mathbf{j+a}\bar{\sigma}}
\delta_{\mathbf{ij+a}}.
\end{eqnarray} 
The last term corresponds to the case of two electrons at the same site
${\bf i} = {\bf j}+{\bf a}$.
It is small and can be omitted. Applying the Hubbard-I decoupling and taking
into account that in the absence of the magnetization
$\langle c_{\mathbf{i}\sigma}c^\dag_{\mathbf{i}\bar{\sigma}}\rangle=0$,
we obtain
\begin{eqnarray}
\label{Hk_fin}
\nonumber
\left[
	H_{\rm kin},
	(n_{\mathbf{i}\bar{\sigma}}+n_{\mathbf{j}\sigma})
	\Delta_{\mathbf{ij}}
\right]
\approx
-t\sum_\mathbf{a}
	\langle n_{\mathbf{i}\bar{\sigma}}\rangle
	c^\dag_{\mathbf{i+a}\sigma}
	c^\dag_{\mathbf{j}\bar{\sigma}}
	\left(1-\delta_{\mathbf{ji+a}}\right)
\\
	+
	\langle n_{\mathbf{j}\sigma}\rangle
	c^\dag_{\mathbf{i}\sigma}
	c^\dag_{\mathbf{j+a}\bar{\sigma}}
	\left(1-\delta_{\mathbf{j+a i}}\right).
\qquad\qquad\qquad
\end{eqnarray}
Then, we calculate the commutator with $H_{\rm int}$: 
\begin{eqnarray}
\left[
	H_{\rm int},
	(n_{\mathbf{i}\bar{\sigma}}+n_{\mathbf{j}\sigma})
	\Delta_{\mathbf{ij}}
\right]
=
(n_{\mathbf{i}\bar{\sigma}}+n_{\mathbf{j}\sigma})
\left[
	H_{\rm int},
	\Delta_{\mathbf{ij}}
\right]
\\
\nonumber
=
U(n_{\mathbf{i}\bar{\sigma}}+n_{\mathbf{j}\sigma})^2
\Delta_{\mathbf{ij}}.
\end{eqnarray} 
We note that
$n_{\mathbf{j}\sigma}^2=n_{\mathbf{j}\sigma}$
and
$
U
n_{\mathbf{i}\bar{\sigma}}
n_{\mathbf{j}\sigma}
\Delta_{\mathbf{ij}}
=
O(t/U)
$.
As a result, it holds:
\begin{eqnarray} 
\left[
	H_{\rm int},
	(n_{\mathbf{i}\bar{\sigma}}+n_{\mathbf{j}\sigma})
	\Delta_{\mathbf{ij}}
\right]
\approx
U(n_{\mathbf{i}\bar{\sigma}}+n_{\mathbf{j}\sigma})
\Delta_{\mathbf{ij}}.
\end{eqnarray} 
Collecting all terms together, we derive
\begin{eqnarray}\label{Hcom}
\nonumber
\left[
	H,(n_{\mathbf{i}\bar{\sigma}}+n_{\mathbf{j}\sigma})
	\Delta_{\mathbf{ij}}
\right]
\approx
-t\langle n_\sigma\rangle
\sum_{\mathbf{a}}
	c^\dag_{\mathbf{i+a}\sigma}
	c^\dag_{\mathbf{j}\bar{\sigma}}
	\left(
		1-\delta_{\mathbf{ji+a}}
	\right)\\
	+
	c^\dag_{\mathbf{i}\sigma}
	c^\dag_{\mathbf{j+a}\bar{\sigma}}
	\left(
		1-\delta_{\mathbf{j+a i}}
	\right)
+
U(n_{\mathbf{i}\bar{\sigma}}+n_{\mathbf{j}\sigma}) \Delta_{\mathbf{ij}},
\qquad\qquad
\end{eqnarray}
where we denote
$\langle
n_{{\bf i}\sigma}\rangle=\langle n_{{\bf j}\bar{\sigma}}\rangle
=\langle n_{\sigma}\rangle$.

Finally, we calculate the commutator:
\begin{eqnarray} 
\left[
	\left(
		n_{\mathbf{i}\bar{\sigma}}
		+
		n_{\mathbf{j}\sigma}
	\right)
	\Delta_{\mathbf{ij}}^{\vphantom{\dagger}},
	\Delta^\dag_{\mathbf{lm}}
\right]
=
\left( 
	n_{{\bf j} \sigma}^{\vphantom{\dagger}}
	c^\dag_{{\bf i} \sigma}
	c^{\vphantom{\dagger}}_{{\bf j} \bar{\sigma}}
	+
	n_{{\bf i} \bar{\sigma}}^{\vphantom{\dagger}} 
	c^\dag_{{\bf i} \sigma}
	c^{\vphantom{\dagger}}_{{\bf l} \sigma}
\right)
\delta_{{\bf j} {\bf m}}
\label{commut}
\\
\nonumber
-
\left(
	n_{{\bf i} \bar{\sigma}}^{\vphantom{\dagger}}
	c^{\vphantom{\dagger}}_{{\bf m} \bar{\sigma}}
	c^\dag_{{\bf j} \bar{\sigma}}
	+
	n_{{\bf j} \sigma}^{\vphantom{\dagger}}
	c_{{\bf m} \bar{\sigma}}^{\vphantom{\dagger}}
	c^\dag_{{\bf j} \bar{\sigma}}
\right)
\delta_{{\bf i} {\bf l}}
\\
\nonumber
-
\left(
	c^{\vphantom{\dagger}}_{{\bf m} \bar{\sigma}}
	c^\dag_{{\bf m} \sigma}
	c^{\vphantom{\dagger}}_{{\bf l} \sigma}
	c^\dag_{{\bf j} \bar{\sigma}}
	\delta_{{\bf i} {\bf m}}
	+
	c^{\vphantom{\dagger}}_{{\bf m} \bar{\sigma}}
	c^\dag_{{\bf i} \sigma}
	c^{\vphantom{\dagger}}_{{\bf l} \sigma}
	c^\dag_{{\bf j} \bar{\sigma}}
	\delta_{{\bf l} {\bf j}}
\right).
\end{eqnarray}
Deriving the latter formula one has to keep in mind that 
$
\delta_{\mathbf{jm}}
\delta_{\mathbf{jl}}
=
\delta_{\mathbf{jm}}
\delta_{\mathbf{lm}}
=0$,
when
$\mathbf{i}\neq \mathbf{j}$
and
$\mathbf{l}\neq \mathbf{m}$.
From (\ref{commut}) it is possible to deduce:
\begin{eqnarray} 
\left\langle\!
	\left[\!
		\left(
			n_{\mathbf{i}\bar{\sigma}}\!
			+\!
			n_{\mathbf{j}\sigma}\!
		\right)
		\Delta_{\mathbf{ij}}^{\vphantom{\dagger}},
		\Delta^\dag_{\mathbf{lm}}
	\right]\!
\right\rangle\!
&\approx &
\left\langle n_{\mathbf{j}\sigma}\right\rangle
\langle 
	c^\dag_{\mathbf{i}\sigma}
	c^{\vphantom{\dagger}}_{\mathbf{l}\sigma}
\rangle
\delta_{\mathbf{jm}}
\label{commut2}
\\
\nonumber
&-&
\left\langle n_{\mathbf{i}\bar{\sigma}}\right\rangle
\langle
	c^{\vphantom{\dagger}}_{\mathbf{m}\bar{\sigma}}
	c^\dag_{\mathbf{j}\bar{\sigma}}
\rangle
\delta_{\mathbf{il}},
\end{eqnarray} 
Obtaining this result we took into account that in the absence of the
magnetic order
\begin{eqnarray} 
\langle
	c^{\vphantom{\dagger}}_{\mathbf{m}\bar{\sigma}}	
	c^\dag_{\mathbf{m}\sigma}
\rangle
=
\langle
	c_{\mathbf{j}\sigma}^{\vphantom{\dagger}} 
	c^\dag_{\mathbf{j}\bar{\sigma}}
\rangle
=0.
\end{eqnarray} 
Further, in (\ref{commut2}) the terms
\begin{eqnarray}
n_{\mathbf{i}\bar{\sigma}}^{\vphantom{\dagger}} 
c^\dag_{\mathbf{i}\sigma}
c^{\vphantom{\dagger}}_{\mathbf{l}\sigma}
\delta_{\mathbf{jm}}
=
O(t/U),
\quad
c_{\mathbf{m}\bar{\sigma}}^{\vphantom{\dagger}}
n_{\mathbf{j}\sigma}^{\vphantom{\dagger}}
c^\dag_{\mathbf{j}\bar{\sigma}}
=O(t/U),
\end{eqnarray}
are dropped. Besides, we performed Hubbard-I decoupling in three-site
operators:
\begin{eqnarray}
n_{\mathbf{j}\sigma}^{\vphantom{\dagger}}
c^\dag_{\mathbf{i}\sigma}
c_{\mathbf{l}\sigma}^{\vphantom{\dagger}}
\approx
\langle n^{\vphantom{\dagger}}_{\mathbf{j}\sigma}\rangle
c^\dag_{\mathbf{i}\sigma}
c_{\mathbf{l}\sigma}^{\vphantom{\dagger}}
(1-\delta_{\mathbf{jl}}),
\label{decoup}
\\
c_{\mathbf{m}\bar{\sigma}}^{\vphantom{\dagger}}
c^\dag_{\mathbf{m}\sigma}
c_{\mathbf{l}\sigma}^{\vphantom{\dagger}}
c^\dag_{\mathbf{j}\bar{\sigma}}
\approx
\langle
	c_{\mathbf{m}\bar{\sigma}}^{\vphantom{\dagger}}
	c^\dag_{\mathbf{m}\sigma}
\rangle
c^{\vphantom{\dagger}}_{\mathbf{l}\sigma}
c^\dag_{\mathbf{j}\bar{\sigma}}
=0.
\end{eqnarray}
In (\ref{decoup}) we used the fact that
$n_{{\bf j} \sigma} c_{{\bf l} \sigma} \delta_{{\bf j} {\bf l}} = 0$.
As a result, within the scope of our approximation,
equation~(\ref{PPP1})
can be written as
\begin{eqnarray*}
\!\!
\dot{{\cal P}}_{\mathbf{ijlm}}\!
\approx\! 
-t
\langle
	n_\sigma
\rangle\!
\sum_{\mathbf{a}}
\left[
	P_{\mathbf{i+a jlm}}
	\left(1\!-\!\delta_{\mathbf{ji+a}}\right)\!
	+\!  
	P_{\mathbf{ij+a lm}}
	\left(1\!-\!\delta_{\mathbf{j+a i}}\right)
\right]
\\
+
\left(U-2\mu\right){\cal P}_{\mathbf{ijlm}}
+
\delta(\tau)
\langle
	n_{\sigma}
\rangle
\left(
	\langle
		c^\dag_{\mathbf{i}\sigma}
		c^{\vphantom{\dagger}}_{\mathbf{l}\sigma}
	\rangle
	\delta_{\mathbf{jm}}
	-
	\langle
		c^{\vphantom{\dagger}}_{\mathbf{m}\bar{\sigma}}
		c^\dag_{\mathbf{j}\bar{\sigma}}
	\rangle
\delta_{\mathbf{il}}\right).
\end{eqnarray*}
Since
${\cal P} = O(t/U)$, 
while
$P = O(1)$,
we omit 
$\dot {\cal P}$
and
$\mu {\cal P}$
terms, arriving at 
equation~(\ref{PPP}).

\section{Fourier transformation}
\label{fourier}

To solve (\ref{PHI1}), analytically or numerically, it is useful to
subject this equation to the Fourier transformation. Let us consider a
square 2D lattice with the number of sites
$N=M\times M$.
In Fourier space
equation~(\ref{PHI1})
reads:
\begin{equation}
\label{Four}
\!\!
P_{\mathbf{k}_1\mathbf{k}_2\mathbf{k}_3\mathbf{k}_4\omega}\!\!
\left[\!
	i\omega\!
	+\!
	\tilde{t}\sum_{\mathbf{a}}
		\left(
			e^{i\mathbf{k}_1\mathbf{a}}\!
			+\!
			e^{i\mathbf{k}_2\mathbf{a}}\!
		\right) \!
		+\!
		2\mu\!
\right]\!
=\!
{\cal S}\!+\!{\cal R}.
\end{equation}
Here
\begin{eqnarray*}
\label{FourSR}
{\cal S}
&=&
\frac{\tilde{t}}{N^2}
\sum_{\mathbf{a ij}}
	e^{i\mathbf{k}_1\mathbf{i}+i\mathbf{k}_2\mathbf{j}}
	[\left(
		\delta_{\mathbf{i+a j}}
		+
		\delta_{\mathbf{ij}}
	\right)
	P_{\mathbf{i+a}\mathbf{j}\mathbf{k}_3\mathbf{k}_4\omega}
\\
	&+&
	\left(
		\delta_{\mathbf{ij+a}}
		+
		\delta_{\mathbf{ij}}
	\right)
	P_{\mathbf{i}\mathbf{j+a}\mathbf{k}_3\mathbf{k}_4\omega}]
\,,
\\
{\cal R}
&=&
\frac{1-\langle n_{\mathbf{i}\sigma}\rangle}{N^2}
\left(
	\langle
		c^\dag_{\mathbf{k}_1\sigma}	
		c_{\mathbf{k}_1\sigma}^{\vphantom{\dagger}} 
	\rangle
	-
	\langle
		c_{\mathbf{k}_2\bar{\sigma}}^{\vphantom{\dagger}} 
		c^\dag_{\mathbf{k}_2\bar{\sigma}}
	\rangle
\right)
\delta_{\mathbf{k}_1\mathbf{k}_3}
\delta_{\mathbf{k}_2\mathbf{k}_4}.
\qquad
\end{eqnarray*}
After straightforward algebra we derive
\begin{equation}
\label{S}
{\cal S}\!\!
= \!\!
\frac{z\tilde{t}}{N}\!
\sum_\mathbf{q}\!  
\left( 
	\gamma_{\mathbf{k}_1}\!\!  
	+\!\!
	\gamma_{\mathbf{k}_2}\!\!  
	+\!\!  
	\gamma_{\mathbf{k}_1-\mathbf{q}}\!\!
	+\!\!  
	\gamma_{\mathbf{k}_2\! +\!\mathbf{q}} 
\right)\!
P_{\mathbf{k}_1-\mathbf{q}, \mathbf{k}_2+\mathbf{q}, \mathbf{k}_3,
\mathbf{k}_4, \omega},
\end{equation}
where the number of the nearest
neighbors on the lattice $z=4$, and we introduce the notation
\begin{equation}
\nonumber
\gamma_{\mathbf{k}}=\frac{1}{z}\sum_{\mathbf{a}}\exp(i\mathbf{ka}).
\end{equation}
The term
${\cal R}$
in the right-hand side of
equation~(\ref{Four})
is transformed with the help of the formula for the single-electron
Green's function in the Hubbard-I approximation:
\begin{eqnarray}
\langle
	c_{\mathbf{i} \sigma}^{\vphantom{\dagger}}
	c^\dagger_{\mathbf{j} \sigma}
\rangle
=
-(1-\langle n_{\bar \sigma} \rangle)
G_0 (\mathbf{i-j}, +0),
\end{eqnarray}
where $G_0$ is the Green's function for free fermions. From this equation
it follows that:
\begin{eqnarray}
\langle 
	c_{\mathbf{k} \sigma}^{\vphantom{\dagger}} 
	c^\dagger_{\mathbf{k} \sigma} 
\rangle 
= 
(1-\langle n_{\bar \sigma} \rangle)
[1-n_{\rm F}(\tilde \varepsilon_\mathbf{k}-\mu)],
\end{eqnarray}
where
$n_{\rm F}$
is Fermi distribution function, and
$\tilde \varepsilon_\mathbf{k}$
is the energy spectrum. In the considered case of the square lattice and
strong electron-electron repulsion
($U\gg zt$)
we can write in the tight-binding approximation:
$\tilde \varepsilon_\mathbf{k}
=
z\tilde t \gamma_{\mathbf{k}}$.

Within the Hubbard-I approach expectation value
$
\langle
	c^\dagger_{\mathbf{j} \sigma}
	c_{\mathbf{i} \sigma}^{\vphantom{\dagger}}
\rangle
$
satisfies:
\begin{eqnarray}
\langle
	c^\dagger_{\mathbf{j} \sigma}
	c_{\mathbf{i} \sigma}^{\vphantom{\dagger}}
\rangle
=
-
\langle
	c_{\mathbf{i} \sigma}^{\vphantom{\dagger}}
	c^\dagger_{\mathbf{j} \sigma}
\rangle
+
(1-\langle n_{\bar \sigma} \rangle) \delta_{\bf ij}.
\end{eqnarray}
This expression is incompatible with the usual Fermi anticommutation
rules. This is an artifact of the Hubbard-I scheme, a discrepancy which is
a consequence of the inexact nature of the Hubbard-I Green's functions.
This means that the Hubbard-I Green's function fails at energy higher than
$\tilde t$, or, equivalently, at
$|{\bf i - j}|$
smaller than at least several lattice constants.

Using the above relations we derive after simple algebra
\begin{equation}\nonumber {\cal
R}\!  =\!  \frac{(1\!-\!\langle n_\sigma\rangle)^2}{M^2}\!  \left[ n_{\rm
F}(\tilde \varepsilon_{\mathbf{k}_1}\!-\!\mu)\!  +\!  n_{\rm F}(\tilde
\varepsilon_{\mathbf{k}_2}\!-\!\mu)\!  -\!1 \right]
\delta_{\mathbf{k}_1\mathbf{k}_3}\delta_{\mathbf{k}_2\mathbf{k}_4}.
\end{equation}
Combining the expressions for ${\cal S}$ and ${\cal R}$, we
can rewrite (\ref{Four}) in the form
\begin{eqnarray}
\label{four1}
\nonumber
\left[
	i\omega
	+
	z\tilde{t}\left(
			\gamma_{\mathbf{k}_1} 
			+ 
			\gamma_{\mathbf{k}_2} 
		  \right) 
	+
	2\mu
\right]
P_{\mathbf{k}_1\mathbf{k}_2\mathbf{k}_3\mathbf{k}_4\omega} 
\\
\nonumber
=
\frac{z\tilde{t}}{M}\!  \sum_{\mathbf{q}}\!  \left(
\gamma_{\mathbf{k}_1}\!  +\!  \gamma_{\mathbf{k}_2}\!  +\!
\gamma_{\mathbf{k}_1-\mathbf{q}}\!  +\!  \gamma_{\mathbf{k}_2-\mathbf{q}}
\right)\!  P_{\mathbf{k}_1-\mathbf{q}, \mathbf{k}_2+\mathbf{q},
\mathbf{k}_3, \mathbf{k}_4, \omega}+ \\ \frac{(1-\langle
n_\sigma\rangle)^2}{M^2}\!  \left[ n_{\rm F}(\tilde
\varepsilon_{\mathbf{k}_1}\!-\!\mu)\!  +\!  n_{\rm F}(\tilde
\varepsilon_{\mathbf{k}_2}\!-\!\mu)\!  -\!1 \right]
\delta_{\mathbf{k}_1\mathbf{k}_3}\delta_{\mathbf{k}_2\mathbf{k}_4}.\quad
\end{eqnarray}
The propagator
$P_{\mathbf{k}_1\mathbf{k}_2\mathbf{k}_3\mathbf{k}_4\omega}$
is non-zero only if
$\mathbf{k}_1+\mathbf{k}_2=\mathbf{k}_3+\mathbf{k}_4$. 
This is a consequence of the momentum conservation law. Therefore, it is
convenient to introduce the total momentum
$\mathbf{k}_1+\mathbf{k}_2=\mathbf{Q}$
and define
\begin{equation}\label{PPt}
P_{\mathbf{k}_i\mathbf{k}_f\omega}^{\mathbf{Q}}= P_{\mathbf{k}_i,
\mathbf{Q}-\mathbf{k}_i, \mathbf{k}_f, \mathbf{Q}-\mathbf{k}_f, \omega}.
\end{equation}
Equation~(\ref{four1})
can be rewritten as
\begin{equation}
\label{FinP}
i\omega
P_{\mathbf{k}_i\mathbf{k}_f\omega}^{\mathbf{Q}}
=
\sum_\mathbf{q}
	h^\mathbf{Q}_{\mathbf{k}_i\mathbf{q}}
	P_{\mathbf{q}\mathbf{k}_f\omega}^{\mathbf{Q}}
	+
	{\cal R}^\mathbf{Q}_{\mathbf{k}_i\mathbf{k}_f},
\end{equation}
where
\begin{eqnarray}
\label{defi}
\nonumber
h^\mathbf{Q}_{\mathbf{k}_i\mathbf{q}}
&=&
-
\left[
	z\tilde{t}
	\left(
		\gamma_{\mathbf{k}_i}
		+
		\gamma_{\mathbf{Q}-\mathbf{k}_i}
	\right)
	+
	2\mu
\right]\delta_{\mathbf{q}\mathbf{k}_i}
\\
&+&
\frac{z\tilde{t}}{M}
\left(
	\gamma_{\mathbf{k}_i}
	+
	\gamma_{\mathbf{Q}-\mathbf{k}_i}
	+
	\gamma_{\mathbf{q}}
	+
	\gamma_{\mathbf{Q}-\mathbf{q}}
\right),
\\
\nonumber
{\cal R}^\mathbf{Q}_{\mathbf{k}_i\mathbf{k}_f}
&=&
\frac{(1-\langle n_\sigma\rangle)^2}{M^2}\!
\\
&\times&
\left[
	n_{\rm F}(\tilde \varepsilon_{\mathbf{k}_i}\!-\!\mu)\!
	+\!
	n_{\rm F}(\tilde \varepsilon_{\mathbf{Q}-\mathbf{k}_i}\!-\!\mu)\!
	-\!1
\right]
\delta_{\mathbf{k}_i\mathbf{k}_f}.
\quad
\end{eqnarray}
Equation~(\ref{FinP})
may be used to calculate the propagator $P$.

\section{Computation of the equal-time correlation function}
\label{comput}

Equation~(\ref{FinP})
may be cast in the matrix form
$i\omega {\mathbf{P}}
=
\mathbf{h}{\mathbf{P}}+\mathbf{\cal R}$.
Matrix
$\mathbf{h}$
is symmetric and, consequently, can be diagonalized. We write down
$\mathbf{h}$
symbolically as
$\mathbf{h}=\mathbf{U}\mathbf{d}\mathbf{U}^+$,
where
$\mathbf{d}$
is a diagonal matrix with the diagonal elements
$d_i$:
\begin{eqnarray}
\mathbf{d} = {\rm diag}(d_i),
\end{eqnarray}
$\mathbf{U}$
is a unitary matrix, $\mathbf{U}^+$ is its Hermitian-conjugated matrix.
Thus, the solution of the problem can be expressed in the form
\begin{equation}\label{Solu}
{\mathbf{P}}(\omega)=
\left[
	\mathbf{U}(i\omega-\mathbf{d})\mathbf{U}^+
\right]^{-1}\mathbf{\cal R}. 
\end{equation}
We diagonalize numerically the matrix $\mathbf{h}$ and
calculate the equal-time propagator
$P_{\mathbf{ijlm}} (+0)
=
\langle
	\Delta_{\mathbf{ij}}^{\vphantom{\dagger}}
	\Delta^\dag_{\mathbf{lm}}
\rangle$
performing frequency summation
\begin{equation}\label{P0}
{\mathbf{P}}(\tau \rightarrow +0)
=
T \sum_{\omega}
	\exp{(i\tau\omega)}\mathbf{P}(\omega)
\Big|_{\tau\rightarrow +0}
\end{equation}
according the rule
\begin{equation*}
T \sum_{\omega}
	\exp{(i\tau\omega)}
	\left(i\omega-\mathbf{d}\right)^{-1}
\Big|_{ \tau\rightarrow +0}
=
N_{\rm B}(\mathbf{d}),
\end{equation*}
where
$N_{\rm B}(\varepsilon)=(e^{\varepsilon/T}-1)^{-1}$
is Bose distribution function, and
$N_{\rm B}(\mathbf{d})$
is a diagonal matrix:
$N_{\rm B}(\mathbf{d})={\rm diag}[N_{\rm B}(d_i)]$. 
As a result, we derive
\begin{equation}\label{Solu1}
{\mathbf{P}}(\tau \rightarrow 0)= {\bf U}
N_{\rm B} (\mathbf{d}){\bf U}^+\mathbf{\cal R}. 
\end{equation}
This equation can be used to calculate the correlation function
numerically.

\end{document}